\documentclass[submission,copyright,creativecommons]{eptcs}

\usepackage[dvipsnames]{xcolor}
\usepackage{listings}
\usepackage{amsmath,amssymb,stmaryrd}
\usepackage{mathpartir}
\usepackage{xspace}
\usepackage{tikz}
\usepackage{wrapfig}
\usepackage{ebproof}
\usepackage{cleveref}
\usepackage{multicol}
\usepackage[nodisplayskipstretch]{setspace}
\usepackage{caption}
\usepackage{rotating}
\usepackage{algorithmicx}
\usepackage[noend]{algpseudocode}
\usepackage[inline]{enumitem}

\crefname{lemma}{lemma}{lemmas}
\Crefname{lemma}{Lemma}{Lemmas}
\crefname{thm}{theorem}{theorems}
\Crefname{thm}{Theorem}{Theorems}

\newcommand{\freest}{\textsc{FreeST}\xspace}

\newcommand{\kindc}[1]{{\color{ForestGreen}{#1}}} 
\newcommand{\typec}[1]{{\color{RoyalBlue}{#1}}}
\newcommand{\termc}[1]{{\color{RedOrange}{#1}}}

\newcommand{\grmeq}{\; ::= \;\;}
\newcommand{\grmor}{\;\mid\;}

\newcommand{\keyword}[1]{\mathsf{#1}}
\newcommand{\syntax}[1]{\mathtt{#1}}

\newcommand{\join}{\sqcup}
\newcommand{\meet}{\sqcap}
\newcommand{\bigjoin}{\bigsqcup}
\newcommand{\subs}[3]{{#3}[{#1}/{#2}]}
\newcommand{\dual}[1]{\overline{#1}}

\newcommand{\subt}{<:}
\newcommand{\subk}{\subt}
\newcommand{\typeofs}{\syntax{typeof}}
\newcommand{\typeof}[1]{\typeofs{(#1)}}
\newcommand{\weakens}{\syntax{Weaken}}
\newcommand{\weaken}[3]{\weakens{(#3,\,#1,\,#2)}}
\newcommand{\merges}{\syntax{Merge}}
\newcommand{\mergeFun}[2]{\merges(#1, #2)}
\newcommand{\multFun}[1]{\syntax{mult}(#1)}

\newcommand{\isClosures}{\syntax{isAbs}}
\newcommand{\isClosure}[3]{\keyword{if}\;\,\isClosures \;#1 \;\,\keyword{then}\;\, #2 \;\,\keyword{else}\;\, #3}

\newcommand{\kindstyle}[1]{\kindc{\textbf{\textsc{#1}}}}
\newcommand{\multstyle}[1]{\kindc{\textbf{\oldstylenums{#1}}}}
\newcommand{\kindmathstyle}[1]{\kindc{#1}} 

\newcommand\mult{\kindc{m}}
\newcommand\lin{\multstyle{1}}
\newcommand\un{\multstyle{*}}
\newcommand\multVar{\kindmathstyle{\varphi}}

\newcommand{\prekind}{\kindmathstyle\upsilon}
\newcommand{\prekindt}{\kindstyle t}
\newcommand{\prekinds}{\kindstyle s}

\newcommand{\kindsym}{\kappa} 
\newcommand{\kind}[1][\kindsym]{\kindmathstyle{#1}}
\newcommand{\kindVar}{\kindmathstyle{\chi}}

\newcommand\kindtu{\un\prekindt}
\newcommand\kindtl{\lin\prekindt}
\newcommand\kindsu{\un\prekinds}
\newcommand\kindsl{\lin\prekinds}
\newcommand\pKind[2]{\kindc{{#1}{#2}}} 

\newcommand\polarity{\typec{\sharp}}
\newcommand\polOut{\typec{!}}
\newcommand\polIn{\typec{?}}

\newcommand\view{\typec{\star}}
\newcommand\viewOut{\typec{\oplus}}
\newcommand\viewIn{\typec{\&}}

\newcommand\rcds[1][\cdot]{\typec{\llparenthesis{#1}\rrparenthesis}}
\newcommand\variant[1][\cdot]{\typec{\langle{#1}\rangle}}
\newcommand\rcd[1][\cdot]{\typec{\{{#1}\}}}
\newcommand\choice[1][\cdot]{\typec{\{{#1}\}}}

\newcommand{\recordm}[4]{\rcds[{#1}{#2}{#3}]_{#4}} 
\newcommand{\record}[4]{\recordm{#1}{#2}{{#3}_{#1}}{{#1}\in{#4}}} 

\newcommand{\recordt}[3]{\record{#1}{\colon}{#2}{#3}} 

\newcommand\type[1][T]{\typec{#1}}

\newcommand\TSkip{\typec{\keyword{Skip}}}
\newcommand\TEnd{\typec{\keyword{End}}}
\newcommand{\TMsg}[2][\polarity]{{#1}\,\typec{#2}}
\newcommand{\TChoice}[2][\view]{{#1}\,\typec{#2}}
\newcommand{\TSemi}[2]{\typec{{#1};{#2}}}
\newcommand{\TUnit}[1][\mult]{\typec{()_{#1}}}

\newcommand{\TFun}[3][\mult]{\typec{{#2} \mathrel{{#1}\!\!\rightarrow}{#3}}}
\newcommand{\TRcdG}[3]{\typec{\recordt{#1}{#2}{#3}}}
\newcommand{\TRcd}[3]{\typec{\rcd[{#1}{\colon}{{#2}_{#1}}]_{{#1}\in{#3}}}}
\newcommand{\TVariant}[3]{\typec{\variant[{#1}{\colon}{{#2}_{#1}}]_{{#1}\in{#3}}}}

\newcommand{\TAll}[3]{\typec{\forall\,{#1}^{#2}\,.\,{#3}}}

\newcommand{\TRec}[3]{\typec{\mu\,{#1}^{#2}\,.\,{#3}}} 

\newcommand{\TVar}[1][a]{\typec{#1}}

\newcommand{\TPair}[2]{\typec{\{\keyword{fst}\colon{#1},\keyword{snd}\colon{#2}\}}}

\newcommand{\recordmExp}[4]{\{{#1}{#2}{#3}\}_{#4}} 
\newcommand{\recordExp}[4]{\recordmExp{#1}{#2}{{#3}_{#1}}{{#1}\in{#4}}} 

\newcommand{\expr}[1][e]{\termc{#1}}

\newcommand\EConst{\expr[c]}
\newcommand\ESend{\expr[\keyword{send}]}
\newcommand\ERcv{\expr[\keyword{receive}]}

\newcommand\EUnit[1][\mult]{\expr[()_{#1}]}

\newcommand\EVal{\expr[v]}

\newcommand\EVar[1][x]{\expr[#1]}

\newcommand\EAbs[4][\mult]{\expr[\lambda_{\kindc{#1}}\,{#2}\colon{\typec{#3}}.\,{\expr[#4]}]}

\newcommand\ETAbs[3]{\expr[\Lambda\,{\typec{#1}}^{\kindc{#2}}.\,{\expr[#3]}]}

\newcommand\ERcd[3]{\expr[\recordExp{#1}{=}{#2}{#3}]}

\newcommand{\ESel}{\expr[\keyword{select}]}

\newcommand\EApp[2]{\expr[#1\;#2]}
\newcommand\ELet[3]{\expr[\keyword{let}\;#1\,=\,#2\;\keyword{in}\;#3]}
\newcommand\ECase[4]{\expr[\keyword{case}\;#1\;\keyword{of}\;{\recordExp{#2}{\rightarrow}{#3}{#4}}]}
\newcommand\squared[1]{[#1]}
\newcommand\ETApp[2]{\expr[#1\,\squared{\typec{#2}}]}
\newcommand{\ENew}{\expr[\keyword{new}]}
\newcommand\EMatch[4]{\expr[\keyword{match}\;#1\;\keyword{with}\;{\recordExp{#2}{\rightarrow}{#3}{#4}}]}

\newcommand{\EPair}[2]{\expr[\{\keyword{fst}={#1},\keyword{snd}={#2}\}]}

\newcommand{\infrule}[3]{\inferrule* [lab=#1]{#2}{#3}}
\newcommand{\axiom}[2]{\infrule {#1}{}{#2}}

\newcommand{\declrel}[2]{\emph{#1}\hfill\fbox{{#2}}}

\newcommand{\twocontexts}[2]{{#1} \mid {#2}}

\newcommand{\judgementrel}[3]{{#1} \; {#2} \; {#3}}

\newcommand{\judgementrelctx}[4]{{#1} \vdash \judgementrel{#2}{#3}{#4}}

\newcommand{\judgementrelbictxinout}[6]{\judgementrelctx{\twocontexts{#1}{#2}}{#3}{#4}{\twocontexts{#5}{#6}}}

\newcommand{\Empty}{\varnothing}

\newcommand{\cset}{\mathcal C} %
\newcommand{\uset}{\Sigma} 

\newcommand{\cGenK}[4][\Delta]{\judgementrelctx{#1}{#2\colon{#3}}{\Rightarrow}{#4}}
\newcommand{\cGenE}[6][\Delta]{\judgementrelbictxinout{#1}{#2}{#3\colon#4}{\Rightarrow}{#5}{#6}}

\newcommand{\rulename}[2]{{#1}-{#2}\xspace}

\newcommand{\rulenameCG}[1]{\rulename{CG}{#1}}
\newcommand\ruleCGSkip{\rulenameCG{Skip}}
\newcommand\ruleCGEnd{\rulenameCG{End}}
\newcommand\ruleCGMsg{\rulenameCG{Msg}}
\newcommand\ruleCGCh{\rulenameCG{Ch}}
\newcommand\ruleCGSeq{\rulenameCG{Seq}}
\newcommand\ruleCGUnit{\rulenameCG{Unit}}
\newcommand\ruleCGArrow{\rulenameCG{Arrow}}
\newcommand\ruleCGRcd{\rulenameCG{Rcd}}

\newcommand\ruleCGTAbs{\rulenameCG{TAbs}}
\newcommand\ruleCGRec{\rulenameCG{Rec}}
\newcommand\ruleCGVar{\rulenameCG{Var}}

\newcommand{\rulenameIE}[1]{\rulename{Inf}{#1}}
\newcommand\rulenameIConst{\rulenameIE{Const}}
\newcommand\rulenameIVar{\rulenameIE{Var}}
\newcommand\rulenameIAbs{\rulenameIE{Abs}}
\newcommand\rulenameIApp{\rulenameIE{App}}
\newcommand\rulenameITAbs{\rulenameIE{TAbs}}
\newcommand\rulenameITApp{\rulenameIE{TApp}}
\newcommand\rulenameIRcd{\rulenameIE{Rcd}}
\newcommand\rulenameIRcdElim{\rulenameIE{RcdElim}}
\newcommand\rulenameIVariant{\rulenameIE{Variant}}
\newcommand\rulenameICase{\rulenameIE{Case}}
\newcommand\rulenameISel{\rulenameIE{Sel}}

\newcommand\rulenameINew{\rulenameIE{New}}

\newcommand{\eg}{e.g.\xspace}  

\newcommand\Small{\small}

\definecolor{darkviolet}{rgb}{0.5,0,0.4}
\definecolor{darkgreen}{rgb}{0,0.4,0.2}
\definecolor{darkblue}{rgb}{0.1,0.1,0.9}
\definecolor{darkgrey}{rgb}{0.5,0.5,0.5}
\definecolor{lightblue}{rgb}{0.4,0.4,1}

\lstdefinestyle{eclipse}{
  breaklines=true,
  basicstyle=\sffamily\Small,
  emphstyle=\color{red}\bfseries,
  keywordstyle=\color{darkviolet}\bfseries,
  commentstyle=\color{darkgreen},
  stringstyle=\color{darkblue},
  numberstyle=\color{darkgrey},
  emphstyle=\color{red},
  showstringspaces=false,
}

\providecommand{\event}{PLACES 2023} 

\usepackage{iftex}

\ifpdf
  \usepackage{underscore}         
  \usepackage[T1]{fontenc}        
\else 
  \usepackage{breakurl}           
\fi

\title{Kind Inference for the FreeST Programming Language
}

\author{
  Bernardo Almeida
  \and
  Andreia Mordido
  \and
  Vasco T. Vasconcelos
  \institute{LASIGE, Faculdade de Ciências, Universidade de Lisboa, Portugal}
  \email{\{bpdalmeida,afmordido,vmvasconcelos\}@ciencias.ulisboa.pt}
}
\def\titlerunning{Kind inference}
\def\authorrunning{B. Almeida, A. Mordido \& V.T. Vasconcelos}
\begin{document}
\maketitle
\begin{abstract}
  We present a kind inference algorithm for the \freest programming language.
  The input to the algorithm is \freest source code with (possibly part of)
  kind annotations replaced by kind variables. The algorithm infers concrete
  kinds for all kind variables.
  We ran the algorithm on the \freest test suite by first replacing kind
  annotation on all type variables by fresh kind variables, and concluded that
  the algorithm correctly infers all kinds. Non surprisingly, we found out that
  programmers do not choose the most general kind in 20\% of the cases.
%
\end{abstract}

\section{Introduction}

Software systems usually handle resources such as files 
and communication channels. 
The correct usage of such resources generally follows a protocol that describes
valid patterns of interactions. For example a file should be opened and
eventually closed, after which no read or write operations should ever be
performed. The case for communication channels is similar: channels are opened,
messages are exchanged, channels may eventually be closed, after which no more
messages should be exchanged.
Session types~\cite{Honda93,HondaVK98,TakeuchiHK94} allow expressing elaborate
protocols (for files and channels, for example) guaranteeing that protocols are
obeyed by programs.
%
%
%

\freest~\cite{AlmeidaMTV22,freest,AlmeidaMV19} is a concurrent
functional programming language based on System F where processes
communicate via heterogeneously typed-channels governed by context-free
session types~\cite{ThiemannV16}.
Context-free session types allow describing protocols such as the serialization
of arithmetic expressions. Consider the following datatype for arithmetic
expressions.
\begin{lstlisting}[numbers=left,xleftmargin=0.6cm]
data Exp = Lit Int | Plus Exp Exp | Times Exp Exp
\end{lstlisting}
An \lstinline|Exp| is either a literal with an integer (\lstinline|Lit Int|), a
sum of two sub-expressions (\lstinline|Plus Exp Exp|) or the product of two
sub-expressions (\lstinline|Times Exp Exp|). To serialise a value of type
\lstinline|Exp| we use a session type such as the following.

\begin{lstlisting}[numbers=left,firstnumber=2,xleftmargin=0.6cm]
type ExpC = +{LitC: !Int, PlusC: ExpC;ExpC, TimesC: ExpC;ExpC}
\end{lstlisting}
The abbreviation \lstinline|ExpC| defines the type of a channel as seen from the
point of view of the writer. A channel of type \lstinline|ExpC| offers a set of
options \lstinline|LitC|, \lstinline|PlusC| and \lstinline|TimesC|. If the first
option is chosen, an integer must be sent (\lstinline|!Int|), while, in the
others, two (sub-) expressions are expected to be sent.

Now, suppose that \lstinline|serialise| is a function that serialises an
\lstinline|Exp| on a channel \lstinline|ExpC|.
%
\begin{lstlisting}[numbers=left,firstnumber=3,xleftmargin=0.6cm]
serialise : Exp -> ExpC;a -> a
\end{lstlisting}
The function expects a channel whose initial part is of type \lstinline|ExpC|
and then behaves as \lstinline|a|: \lstinline|serialise| is thus polymorphic on
\lstinline|a|. It consumes the front of the channel (of type \lstinline|ExpC|)
and returns the unused part of the channel (of type \lstinline|a|).

As simple as it may seem, the above code is not valid in the current version of
\freest. The actual code requires further annotations allowing to distinguish
functional from session types as well as linear from unrestricted types.
The distinction is materialised by classifying types with kinds.

In \freest kinds are composed of a multiplicity and a basic kind. Multiplicities
control the number of times a value may be used: exactly once (linear, 
\lstinline|1|) or zero or more (unrestricted, \lstinline|*|).
Basic kinds distinguish functional types (\lstinline|T|) from session types (\lstinline|S|).
The reason why \freest requires kinds lies on polymorphism. If
\lstinline|!Int;?Int| is undoubtedly a session type and \lstinline|Int -> Bool|
a functional type, the same does not apply to the polymorphic variable
\lstinline|a|. Is it a session type or a functional type? The answer depends on
the base kind of \lstinline|a|: if \lstinline|S| or then it is a session type,
if \lstinline|T| then it is a functional type. Kinds are thus necessary to
decide whether the types such as \lstinline|a;!Int| are well-formed.

The datatype defined in line 1 is currently written in annotated form as follows.
\begin{lstlisting}[numbers=left,firstnumber=4,xleftmargin=0.6cm]
data Exp:*T = Lit Int | Plus Exp Exp | Times Exp Exp
\end{lstlisting}
The kind annotation \lstinline|*T|, says that the datatype is functional. As for
the multiplicity, we chose the unrestricted usage so that it may be used as
often as required. Notwithstanding, one may declare \lstinline|Exp| of kind
\lstinline|1T|, in which case \lstinline|serialise| must become a linear function
(of type \lstinline|Exp -> ExpC;a 1-> a|).

Expanding the abbreviation and annotating the datatype in line 2 we get the
following type.
\begin{lstlisting}[numbers=left,firstnumber=5,xleftmargin=0.6cm]
type ExpC:1S = rec a:1S . +{LitC: !Int, PlusC: a;a, TimesC: a;a}
\end{lstlisting}
\lstinline|ExpC| defines a recursive type that is well-formed when the kind of
its body, the external choice ($\oplus$), is a subkind of the kind for the
recursion variable. In this case, the recursion variable \lstinline|ExpC| is
annotated with \lstinline|1S|, given that its body is itself a linear session.

Finally, the function \lstinline|serialise| is currently written as follows.
\begin{lstlisting}[numbers=left,firstnumber=6,xleftmargin=0.6cm]
serialise : foralla:1S . Exp -> ExpC;a -> a
\end{lstlisting}
The polymorphic variable \lstinline|a| stands for the continuation channel;
it must be a linear session. Annotating \lstinline|a| with the unrestricted
session \lstinline|*S| would dictate that it can only be instantiated with
$\TSkip$, the only unrestricted session type.


Even if kinds are necessary in the underlying theory of the \freest language,
they clutter the code.
The code in lines 1--3 is easier to understand and quicker to write; programmers
need not fight the subtleties of each kind. Note that once kinds are inferred,
the prenex occurrences of $\forall$ can be omitted.
The algorithm that we present in this paper annotates all type variables with
their kinds, converting the code in lines 1--3 to that in lines 4--6.

The works more closely related to \freest are Quill~\cite{Morris16},
Affe~\cite{RadanneST20}, Alms~\cite{TovP11},
$\text{F}^{\,\circ}$~\cite{MazurakZZ10},
$\text{FuSe}^{\{\}}$~\cite{Padovani19} and Linear
Haskell~\cite{BernardyBNJS18}. All these languages feature substructural
type systems for dealing with linear, functional and affine types (in the case
of Affe).


Quill~\cite{Morris16} is a language with linear types and a syntax similar to
that of Haskell. Quill features a novel design that combines linear and
functional types. Contrarily to \freest, Quill does not use
kind mechanisms to distinguish between linear and functional types, instead it
uses type predicates (or, qualified types) to reason about linearity.
Furthermore, Quill does not support subkinding. Quill also has a type inference
algorithm which was proven sound and complete.
Affe~\cite{RadanneST20} is an ML-like language with support to linear, affine
and unrestricted types. Like Quill, Affe uses kinds and constrained types to
distinguish between linear and affine types. Affe supports subkinding and it is
equipped with full principal type inference.
Like Affe, Alms~\cite{TovP11} is an ML-like language but is based on
$\text{System F}_{<:}^\omega$, the higher-order polymorphic $\lambda$-calculus
with subtyping. Alms supports affine and unrestricted types. It features a rich
kind system with dependent kinds, unions, and intersections. Moreover, Alms supports ML
modules, allows to expose unrestricted types as affine which gives
flexibility to library programmers and it is equipped with local type
inference. 
$\text{F}^{\,\circ}$~\cite{MazurakZZ10} is an extension of System F that
uses kinds to distinguish between linear and unrestricted types. Similarly to
Affe and Alms, it supports subkinding. Similarly to \freest, but unlike Affe,
$\text{F}^{\,\circ}$ does not support quantification over kinds.
%
%
The work closest to \freest in terms of context-free session types is
$\text{FuSe}^{\{\}}$~\cite{Padovani19}. Padovani proposed an alternative
formulation of context-free session types in which code and types are aligned
via extra annotations, something we decided to avoid in \freest.
Linear Haskell~\cite{BernardyBNJS18} is a proposal to bring linear types to
Haskell. In Linear Haskell functions $T \rightarrow U$ and $T \multimap U$
describe how the arguments of the function are used. The latter form, inspired
by linear logic~\cite{Girard87}, uses the argument $T$ exactly once. In \freest,
annotated arrows $\TFun[\un]TU$ or $\TFun[\lin]TU$ describe how the function is
used (unbounded usage or exactly once).
\freest kinding system differentiates session from functional types. It also
classifies types according to their usage, linear or unrestricted. Other systems
consider these notions separately (or only one of them). The ideas behind our
inference algorithm are similar to Quill and Affe, but the details are quite
different since we do not use type qualifiers to reason about linearity.




\section{The Syntax of Kinds, Types and Expressions}
\label{sec:kinds-types}

This section briefly introduces the notions of kinds, types and expressions; we
refer the interested reader to previous work for details~\cite{AlmeidaMTV22}.
\freest relies on a base set for type variables (denoted by $\typec a$,
$\typec b$, $\typec c$) and another for labels (denoted by $\typec k$,
$\typec \ell$). For the purpose of kind inference, we further use
\emph{multiplicity variables} (denoted by $\multVar$) and \emph{kind variables}
(denoted by $\kindVar$).
The syntax of kinds, types and expressions is in \cref{fig:types}.

\begin{figure}[t]
  \begin{align*}
    \mult     \grmeq& \un \grmor \lin \grmor \multVar             &\text{Multiplicity}\\
    \prekind  \grmeq& \prekinds \grmor \prekindt  &\text{Prekind}\\
    \kind     \grmeq& \mult \prekind \grmor \kindVar  &\text{Kind}\\
    \polarity \grmeq& \polOut \grmor \polIn       &\text{Polarity}\\
    \view     \grmeq& \viewOut \grmor \viewIn     &\text{View}\\
    \rcds     \grmeq& \rcd \grmor \variant &\text{Record}\\
    \type     \grmeq& \TSkip
                      \grmor \TEnd
                      \grmor \TMsg T
                      \grmor \TChoice{\recordt\ell TL}
                      \grmor \TSemi TT            
                      \grmor \TUnit &\text{Type}\\
              \grmor& \TFun TT
                      \grmor \TRcdG\ell TL 
                      \grmor \TAll a\kind T 
                      \grmor \TRec a\kind T
                      \grmor \TVar         &\\
    \expr
    \grmeq& \EUnit
            \grmor \EVar 
            \grmor \EAbs{x}{\type}{\expr}
            \grmor \ETAbs\TVar\kind\EVal
            \grmor \EApp\expr\expr           
            \grmor \ERcd \ell \expr L
            \grmor \ELet{\ERcd \ell x L}\expr\expr &\text{Expression}\\
    \grmor& \EApp \ell \expr 
            \grmor \ELet\EUnit\expr\expr
            \grmor \ECase\expr \ell x L
            \grmor \ETApp \expr \type
            \grmor \EMatch\expr \ell x L
  \end{align*}
  \caption{The syntax of kinds and types with support for kind inference}
  \label{fig:types}
\end{figure}


\begin{wrapfigure}{r}{0.15\textwidth}
  \begin{tikzpicture}[scale=.60]
    \node (TL) at (0,1.5) {$\kindtl$};
    \node (TU) at (-1.5,0) {$\kindtu$};
    \node (SL) at (1.5,0) {$\kindsl$};
    \node (SU) at (0,-1.5) {$\kindsu$};
    \draw (TL) -- (TU) -- (SU) -- (SL) -- (TL);
    \draw (SL) -- (TL);
    \draw (SU) -- (TU);
  \end{tikzpicture}
\end{wrapfigure}
\emph{Multiplicities} are used to indicate the number of times a value can be
used. They are either unrestricted ($\un$), which denotes an arbitrary number of
usages, linear ($\lin$), indicating precisely one usage, or a multiplicity
variable ($\multVar$). The kinding system relies on two base kinds: $\prekinds$
for session types and $\prekindt$ for arbitrary types. \emph{Kinds} are either
the combination of a base kind and a multiplicity or a kind variable $\kindVar$.
%
Since a value of an unrestricted type may be used zero or more times, and
one with a linear type must be used exactly once, it should be clear that
an unrestricted value can used where a linear one is expected. Similarly, the
interpretation of base kinds should be such that a session type ($\kindsu$,
$\kindsl$) can be used in place of an arbitrary type ($\kindtl$). The subkind
relation for non variables (denoted $\kind \subk \kind$) forms a lattice, as
exhibited in the diagram.

\emph{Session types} include $\TSkip$ indicating no communication, $\TEnd$
representing channels ready to be closed, output ($\TMsg[\polOut] T$) and input
($\TMsg[\polIn] T$) messages, internal
($\TChoice[\viewIn]{\choice[{\ell}{\colon}{T_\ell}]_{\ell\in L}}$) and external
choices ($\TChoice[\viewOut]{\choice[{\ell}{\colon}{T_\ell}]_{\ell\in L}}$) and
sequential composition ($\TSemi TU$).
\emph{Functional types} are composed of linear $\TUnit[\lin]$ and unrestricted unit
types $\TUnit[\un]$, linear $\TFun[\lin]TU$ and unrestricted $\TFun[\un]TU$
functions, records $\TRcd\ell TL$, variants $\TVariant\ell TL$ and universal
types $\TAll a\kind T $.
Recursive types $\TRec a \kind T$ are either session or functional depending on
$\kind$. Type variables $\TVar$ may refer to recursion variables in recursive
types or to polymorphic variables in universal types.
%
A function capturing in its body a free linear variable must itself be
linear.

\begin{sloppypar}
  \emph{Expressions} include variables $\EVar$,
  term abstraction $\EAbs{x}{\type}{\expr}$ and application $\EApp ee$,
  type abstraction $\ETAbs\TVar\kind\EVal$ and application $\ETApp\expr\type$,
  record $\ERcd \ell \expr L$ and record elimination
  $\ELet{\ERcd \ell x L}\expr\expr$,
  unit $\EUnit$ and unit elimination $\ELet\EUnit\expr\expr$,
  injection in a variant $\EApp \ell \expr$ and variant elimination
  $\ECase\expr \ell x L$. The expressions for channel operations include channel
  creation, $\EApp\ENew\type$, and branching on a choice,
  $\EMatch\expr \ell x L$. The remaining operations on channels---namely
  $\ENew$, $\ESend$, $\ERcv$ and $\EApp \ESel \ell$---are all understood as
  constants (pre-defined variables).
\end{sloppypar}
Given that our goal is to infer kind annotations, the reader may wonder why we
allow them in the source language, namely in polymorphic types
$\TAll a\kind T $, in recursive types $\TRec a \kind T$ and in type abstractions
$\ETAbs\TVar\kind\EVal$. Programmers may, if they so wish, provide kind
annotations in the source code. Such annotations are passed to the algorithm.
For those omitted, a fresh kind variable $\kindVar$ is generated in its place.





\section{Kind Inference}
\label{sec:kindinference}

Our approach to kind inference follows the established two-step process, wherein
the first gathers constraints and the second resolves the constraints.
The constraint generation step produces constraints in two forms:
$\kind\subk\kind$ and
$\multVar = \bigjoin_{\ell\in L}\multFun{\kind[\kindsym_\ell]}$.
The first form represents subkinding constraints, while the second 
represents equalities between multiplicity variables and the least upper bound
of a given set of multiplicities. To enhance readability, we use shorthand
notation $\multVar = \multFun{\kind[\kind]}$ for
$\multVar = \bigjoin\multFun{\kind[\kind]}$ and use $\bigjoin$ in infix format
for binary sets.




\paragraph{Constraint Generation from Types}

Kind and multiplicity constraints are captured by judgement
$\cGenK[\Delta_{\text{in}}]{\type_{\text{in}}}{\kind_{\text{out}}}{\cset_{\text{out}}}$.
The judgement states that type $\type$ has kind $\kind$ under kinding context
$\Delta$ (a map from type variables to kinds), producing constraint set $\cset$.
To clarify the distinction between input and output, we use the subscript
\textbf{in} for parameters and \textbf{out} for results.


%
%

We explain a core subset of the constraint generation rules, those in
\cref{fig:type-contraints} (the complete set is in
\cref{fig:type-contraints-full}).
Rule \ruleCGVar reads the kind for type variable $\TVar$ (recursive or
polymorphic) from the kinding context, generating no additional restrictions.
Rule \ruleCGRec governs recursive types which can either be session or
functional. The kind of the recursion variable is copied to the kinding context when
analysing type $\typec T$. A constraint $\kindc{\kind'} \subk \kind$ is
generated to ensure that the kind $\kindc{\kind'}$ of the body of the
recursive type is a subkind of the kind $\kind$ of the recursion variable.
Rule \ruleCGArrow, deals with functions $\TFun{\type}{\type[U]}$. It applies the
algorithm recursively to $\type$ and $\type[U]$, and assigns the kind
$\pKind\mult\prekindt$ to the function type, where $\mult$ comes from the arrow
annotation.
Rule \ruleCGRcd builds kinds and constraints for all elements in the record. It
generates a new fresh multiplicity variable $\multVar$. The result is kind
$\pKind\multVar\prekindt$ and the constraint set is composed of the union of
$\cset_\ell$ for all $\ell\in L$ and a new constraint associating variable
$\multVar$ to the least upper bound of the multiplicities of
$\kind[\kind_\ell]$. In order to ensure that $\multVar$ gets the expected
multiplicity, all elements must be subkinds of the kind of the record, that is
$\pKind\multVar\prekindt$. Thus, if at least one entry in the record is linear,
then $\multVar$ is also constrained to be linear.
Rule \ruleCGTAbs adds the kind of the polymorphic variable to the typing context
when checking the body $\typec T$. It then assigns kind
$\kind[\pKind\multVar\prekindt]$ to the incoming type $\TAll a\kind\type$, where
the fresh multiplicity variable $\multVar$ denotes the multiplicity of the kind
of type $\type$.

\begin{figure}[t] 
  \declrel{}{$\cGenK[\Delta_{\text{in}}]{\type_{\text{in}}}{\kind_{\text{out}}}{\cset_{\text{out}}}$}
  \begin{gather*}
      \axiom{\ruleCGVar}
            {\cGenK[\Delta, \TVar\colon\kind] \TVar\kind\Empty}
    \quad        
      \infrule{\ruleCGRec}
              {\cGenK [\Delta, \TVar\colon\kind] \type {\kindc{\kind'}} \cset}
              {\cGenK {\TRec a\kind\type} {\kindc{\kind'}}{\cset \cup \{{\kindc{\kind'}} \subk \kind\}}}
     \quad
      \infrule{\ruleCGArrow}
              {\cGenK \type{\kind[\kindsym_1]}{\cset_1} \\
               \cGenK {\type[U]}{\kind[\kindsym_2]}{\cset_2}}
              {\cGenK{\TFun{\type}{\type[U]}}{\pKind\mult\prekindt}{\cset_1 \cup \cset_2}}
    \\        
            \infrule{\ruleCGRcd}
              {\cGenK{\type[T_\ell]}{\kind[\kindsym_\ell]}{\cset_\ell} \\
                \multVar\;\text{fresh}\\
                (\forall \ell\in L)
              }
              {\cGenK{\TRcd\ell TL}{\pKind\multVar\prekindt}
                {\bigcup_{\ell\in L}  \cset_\ell \cup
                  \{\multVar = \bigjoin_{\ell\in L}\multFun{\kind[\kindsym_\ell]}
                  , \kind[\kind_\ell] \subk \pKind\multVar\prekindt
                  \}}}
      \infrule{\ruleCGTAbs}
              {\cGenK[\Delta, \TVar\colon\kind] \type{\kindc{\kind'}}\cset \\\multVar\;\text{fresh}}
              {\cGenK {\TAll a\kind\type} {\pKind\multVar\prekindt} {\cset \cup \{\multVar = \multFun{\kindc{\kind'}}\}}}
  \end{gather*}
  \vspace*{-8mm}
  \caption{Constraint generation from types}
  \label{fig:type-contraints}
\end{figure}


Type operator $\syntax{mult}$ is fully resolved only after analysing
expressions. At this point it can only be partially resolved. When applied to a
kind of the form $\pKind\mult\prekind$ operator $\syntax{mult}$ rewrites into
multiplicity $\mult$, that is, $\multFun{\pKind\mult\prekind} = \mult$.


As an example, let us consider the function that extracts the first element of a pair.
\begin{equation*}
  \syntax{fst}\colon\TAll a{\kindc{\kindVar_a}}{\TAll b{\kindc{\kindVar_b}}
  {\TFun[\un]{\TPair{\TVar}{\TVar[b]}}{\TVar}}}
\end{equation*}
\sloppy The application of the rules in \cref{fig:type-contraints}, yields the
constraint set
$\{
\kind[\multVar_1] = \multFun{\pKind{\multVar_2}\prekindt},
\kind[\multVar_2] = \multFun\kindtu,
\kind[\multVar_3] = \multFun{\kind[\kindVar_a]} \join
\multFun{\kind[\kindVar_b]}
\}$. Solving the
constraint set one obtains
$\{
\kind[\multVar_1] = \un,
\kind[\multVar_2] = \un,
\kind[\multVar_3] = \multFun{\kind[\kindVar_a]} \join
\multFun{\kind[\kindVar_b]}
\}$. We resolve the indeterminacy of kind variables $\kind[\kindVar_a]$ and
$\kind[\kindVar_b]$ by assuming that they both are $\kindtl$, the maximum of
the kind lattice. 
The solution would then be
$\{\kind[\multVar_1] = \un$, $\kind[\multVar_2] = \un$,
$\kind[\multVar_3] = \lin$,
$\kindc{\kindVar_a} = \kindtl, \kindc{\kindVar_b} = \kindtl\}$.

We argue that assigning $\kindtl$ (the maximum) to $\kindc{\kindVar_a}$ and
$\kindc{\kindVar_b}$ is the preferred solution, since it is the less restrictive
of all solutions. If we were to choose another kind, such as $\kindtu$, then it
would be impossible to call function $ \syntax{fst}$ on linear values (of types
with kind $\kindtl$). We would, undesirably, be ruling out some perfectly
well-behaved programs.

But is $\kindtl$ the best kind for variables $\kindc{\kindVar_a}$ and
$\kindc{\kindVar_b}$? The answer depends on the definition of $\syntax{fst}$.
\begin{equation*}
\syntax{fst} = \ETAbs \TVar {\kindVar_a} {\ETAbs {\TVar[b]} {\kindVar_b}
  {{\EAbs[\un] {p} {\TPair{\TVar}{\TVar[b]}} {\ELet{\EPair xy}{p}{x}}}}}  
\end{equation*}

An examination of expression $\ELet{\EPair xy}{p}{x}$ reveals that the second
element of the pair, $\termc y$, is discarded. Hence, $\kind[\kindVar_b]$ must
be unrestricted. Would $\kind[\kindVar_b]=\kindtl$ be chosen, then \freest would
complain about a linearity violation when type checking the function. In other
words, constraint $\kind[\kindVar_b] \subk \kindtu$ must be added to the
constraint set, but an inspection of the type of $\syntax{fst}$ alone does not
provide enough information to generate such a constraint. In the following, we
present rules that allow generating constraints such as
$\kind[\kindVar_b] \subk \kindtu$ by inspecting variable usage in expressions.


\paragraph{Constraint Generation from Expressions}

\begin{figure}[t] 
  \declrel{}{$\cGenE[\Delta_{\text{in}}]{\Gamma_{\text{in}}}{\expr_{\text{in}}}{\type_{\text{out}}}{\cset_{\text{out}}}{\uset_{\text{out}}}$}
  \begin{gather*}
    \infrule{\rulenameIVar}
            {\cGenK\type\kind\cset}
            {\cGenE{\Gamma,\EVar\colon\type}\EVar\type\cset{\{\EVar\colon\kind\}}}
    \\
            \quad
    \infrule{\rulenameIAbs}
            {\cGenK{\type[T_1]}\kind\cset_1 \quad
              \cGenE{\Gamma,\EVar\colon\type[T_1]}\expr{\type[T_2]}{\cset_2}\uset
              \quad
              \cset_3 = \isClosure{\expr}{\{\kind\subk{\pKind\mult\prekindt}\}}{\Empty}
            }
            {\cGenE\Gamma{\EAbs x{\type[T_1]}\expr} {\TFun{T_1}{T_2}}
              {\cset_1 \cup \cset_2\cup \cset_3 \cup \weaken{\EVar}{\kind}{\uset}}
              {\uset\setminus\EVar}}
            \\
    \infrule{\rulenameIApp}
            {\cGenE\Gamma{\expr[e_1]}{\TFun{T_1}{T_2}}{\cset_1}{\uset_1}
              \\
              \cGenE\Gamma{\expr[e_2]}{\type[T_1]}{\cset_2}{\uset_2}
              \\
              \cGenK{\TFun{T_1}{T_2}}\kind{\cset_3}
            }
            {\cGenE\Gamma{\EApp {e_1}{e_2}}{\type[T_2]}
              {{\cset_1}\cup{\cset_2}\cup{\cset_3}\cup{\mergeFun{\uset_1}{\uset_2}}}
              {\uset_1 \cup \uset_2}}
            \\
   \infrule{\rulenameIRcdElim}
            {\cGenE\Gamma{\expr[e_1]}{\TRcd\ell TL}{\cset_1}{\uset_1}
              \\
              \cGenE{\Gamma,(\expr[x_\ell]\colon{\type[T_\ell]})_{\ell\in L}}
                {\expr[e_2]}\type{\cset_2}{\uset_2}
              \\
              \cGenK{\type}\kind{\cset_3}
              \\
              \cGenK{\type[T_\ell]}{\kind[\kindsym_\ell]}{\cset_\ell}
              \\
              \cset = \cset_1 \cup \cset_2 \cup \cset_3 \cup \cset_\ell \cup \mergeFun{\uset_1}{\uset_2} \cup
              \weaken{\expr[x_\ell]}{\kind[\kindsym_\ell]}{\uset_2}
              \\(\forall \ell\in L)
            }
            {\cGenE\Gamma{\ELet{\ERcd \ell x L}{e_1}{e_2}}\type
              \cset{(\uset_1\cup\uset_2)}\setminus\{ \expr[x_\ell] \}_{\ell\in L} }
  \end{gather*}
  \caption{Constraint generation from expressions}
  \label{fig:exp-cgen}
\end{figure}


Constraints for expressions are derived from judgement
$\cGenE[\Delta_{\text{in}}] {\Gamma_{\text{in}}} {\expr_{\text{in}}}
{\type_{\text{out}}} {\cset_{\text{out}}} {\uset_{\text{out}}}$. The judgement
states that expression $\expr$ has type $\type$ under kinding context $\Delta$
and typing context $\Gamma$. It generates a constraint set $\cset$ and a usage
context $\uset$. Typing contexts map term variables $\termc x$ to types
$\typec T$; usage contexts map term variables $\EVar$ to the kind $\kind$ of
their types. Usage contexts enable reasoning about variable usage: if the
variable is used exactly once, it may be linear, otherwise it must be
unrestricted. Next, we define functions $\weakens$ and $\merges$. The former
checks whether variables are used in expressions. If a variable is not used,
then the set with constraint $\kind\subk\kindtu$ is returned. The latter checks
whether a variable is used more than once: if it appears in multiple usage
contexts, it must also be unrestricted.
\vspace*{-2em}
\begin{multicols}{2}
    \begin{equation*}
    \weaken {\expr[x]}\kind\uset =
    \begin{cases}
      \Empty                & \text{if}\, \expr[x] \in\uset \\
      \{\kind\subk\kindtu\} & \text{otherwise}
    \end{cases}       
  \end{equation*}
  \break 
  \begin{equation*}
    \mergeFun{\uset_1}{\uset_2} = \{\kind\subk\kindtu \mid \expr[x]\colon\kind\in\uset_1 \cap \uset_2\}    
  \end{equation*}
\end{multicols}


%
We are now in a position to explain the rules for expressions,
in~\cref{fig:exp-cgen} (the complete set is in \cref{fig:exp-cgen-full}).
Rule \rulenameIVar is used to assign a type to a variable in a given typing
context. The rule requires the type context $\Gamma$ to contain an entry
$\EVar\colon\type$. The constraints pertaining to type $\type$ are gathered in
$\cset$. To reflect the usage of $\EVar$, the rule returns a singleton map
$\EVar\colon\kind$, where $\kind$ is the kind of $\type$.
Rule \rulenameIAbs deals with abstractions $\EAbs x{\type[T_1]}\expr$. It
recursively calls the judgments on $\type[\type_1]$ and on $\expr$ to collect
constraint sets $\cset_1$, $\cset_2$ and usage context $\uset$. The rule
uses a new predicate, $\isClosures\;\expr$, which holds when $\expr$ is an
abstraction. Then, if $\expr$ is a closure the kind of $\type[\type_1]$ must be
a subkind of $\pKind\mult\prekindt$, where $\mult$ is the multiplicity of the
abstraction. This restriction ensures that unrestricted abstractions do not
close over linear values. The result is type $\TFun{T_1}{T_2}$ together with a
constraint set composed of the union of $\cset_1$, $\cset_2$, $\cset_3$ and the
result of $\weakens$.
The $\weakens$ function checks whether a variable is unused at the end of its
scope. In this case, the lambda abstraction introduces term variable $\EVar$ and
therefore, at the end of the scope, we have to check its usage.
Rule \rulenameIApp states that if $\expr[\expr_1]$ has type $\TFun{T_1}{T_2}$
and $\expr[\expr_2]$ has type $\type[\type_1]$, then the expression
$\EApp{\expr_1}{\expr_2}$ has type $\type[\type_2]$. The constraints $\cset$ and
usage context $\uset$ are computed by combining the results of the kind
inference of $\expr[\expr_1]$, $\expr[\expr_2]$ and $\type$. The final
constraint set is the union of $\uset_1$, $\uset_2$, $\uset_3$, and the result
of the $\merges$ function which imposes that any variable found in both
$\uset_1$ and $\uset_2$ must be unrestricted. The final usage
context is $\uset_1 \cup \uset_2$.
Rule \rulenameIRcdElim combines all previously discussed concepts: it evaluates
expressions $\expr[\expr_1]$ and $\expr[\expr_2]$, collecting $\cset_1, \cset_2$
and $\uset_1, \uset_2$. The result is the type of $\expr[\expr_2]$, a constraint
set $\cset$, which is the union of $\cset_1, \cset_2, \cset_3$, the result
of $\merges$ on $\uset_1$ and $\uset_2$,
and the application of $\weakens$ on $\uset_2$ for all
$\expr[x_\ell]\colon\kind[\kindsym_\ell]$ to check for unused variables. 
The resulting usage context is the combination of $\uset_1$ and $\uset_2$ with all
entries for $\expr[x_\ell]$ removed.

When analysing constraint generation from the type for function $\syntax{fst}$,
we intuitively concluded that the second element in the pair must be
unrestricted because it is discarded. The application of rules in
\cref{fig:exp-cgen}, yield the constraint set
$\{\kindc{\kindVar_b} \subk \kindtu, \kindc{\kindVar_a} \subk
\pKind{\multVar_1}\prekindt, \kindc{\kindVar_b} \subk
\pKind{\multVar_1}\prekindt, \kindc{\kindVar_a} \subk
\pKind{\multVar_0}\prekindt, \kindc{\kindVar_b} \subk
\pKind{\multVar_0}\prekindt, \kindc{\multVar_0} = \multFun{\kindc{\kindVar_a}}
\join \multFun{\kindc{\kindVar_b}}, \kindc{\multVar_1} =
\multFun{\kindc{\kindVar_a}} \join \multFun{\kindc{\kindVar_b}}\}$. A solution
for this set is
$\{\kindc{\multVar_0} = \lin, \kindc{\multVar_1} = \lin, \kindc{\kindVar_a} =
\kindtl, \kindc{\kindVar_b} = \kindtu\}$. The kind variable $\kindc{\kindVar_b}$
is set to $\kindtu$ as we predicted.
The constraint set is computed by combining the constraint sets generated
resulting from applying the judgement to all sub-expressions and the result of
functions $\merges$ and $\weakens$. First, we examine the $\merges$ function: it
takes contexts ${\{{\EVar[p]}\colon{\kind[\kind_p]}\}}$ and
${\{{\EVar[x]}\colon{\kind[\kindVar_a]}\}}$ as input and calculates the
intersection of the two contexts, adding a constraint $\kind\subk\kindtu$ for
each element in the intersection. This process ensures that any variable that is
used in both contexts is unrestricted.
The $\weakens$ function is used to verify if any newly introduced variable is
eventually discarded. In our example, $\weakens$ is applied to
$\EVar\colon{\kind[\kindVar_a]}$ and ${\EVar[y]}\colon{\kind[\kindVar_b]}$
against usage context
$\{{\EVar[p]}\colon{\kind[\kind_p]},\, {\EVar[x]}\colon{\kind[\kindVar_a]}\}$.
For ${\EVar[y]}\colon{\kind[\kindVar_b]}$ function $\weakens$ proceeds as
follows: since $\termc y$ is not present in the context, a new constraint
$\{\kind[\kindVar_b]\subk\kindtu\}$ is added. On the other hand, since $\EVar$
is already in the context, no constraint is created.

\paragraph{Constraint Solving}

We now describe an algorithm to solve constraint sets.

\begin{enumerate}
\item Initialise all kind variables $\kindVar$ to the maximum of the kind
  lattice, $\kindtl$. Likewise initialize all multiplicity variables $\multVar$
  to the maximum of multiplicities, $\lin$. Store them in $\sigma$.
  
\item Iterate over each constraint in the set:
  \begin{enumerate}

  \item If the constraint is of the form $\kind[\kindVar]\subk\kind$, then
    update the entry for $\kindVar$ in $\sigma$ with the greatest lower bound of
    $\kind$ and $\sigma(\kindVar)$.
    For example, if $\sigma = [\kindVar\mapsto\kindtl]$ and we are analysing
    constraint $\kindVar\subk\kindtu$, then the value for $\kindVar$ in $\sigma$
    must be updated to $\kindtl\meet\kindtu = \kindtu$. After this step, we
    would have $\sigma = [\kindVar\mapsto\kindtu]$.
    
  \item If the constraint is of the form $\kind\subk\kindVar$, then check whether
    $\kind$ and the kind for $\kindVar$ in $\sigma$ is in the subkind relation;
    if not then fail. For example, if $\sigma = [\kindVar\mapsto\kindtl]$ and we are
    analysing constraint $\kindtu\subk\kindVar$, then we find that it is in the
    subkind relation since $\kindtu\subk\kindtl$. A failure would happen with
    $\sigma = [\kindVar\mapsto\kindtl]$.



  \item If the constraint is of the form $\kind[\kind_1]\subk\kind[\kind_2]$ and
    neither of the elements is a kind variable, then check whether
    $\kind[\kind_1]\subk\kind[\kind_2]$ is in the subkind relation; if not then
    fail. If not fail, then remove constraint $\kind[\kind_1]\subk\kind[\kind_2]$
    from the constraint set.
    
  \item If the constraint is a multiplicity constraint
    $\multVar = \bigjoin_{\ell\in L}\multFun{\kind[\kindsym_\ell]}$, then compute the
    least upper bound of the multiplicities. If any $\kind[\kindsym_\ell]$ is a
    kind variable ($\kindVar$) or a base kind with a multiplicity variable
    ($\pKind\multVar\prekindt$), we get its kind from $\sigma$ (recall that all
    variables are in $\sigma$ as per step 1).
    If the thus obtained kind is more restrictive than that for $\multVar$
    in $\sigma$ (\eg $\un$ against $\sigma(\multVar) = \lin$), then store it in
    $\sigma$. If $\multVar = \un$, then remove the constraint from the set.
  \end{enumerate}
\item Repeat the process in step 2 until there are no further updates to be made.
\item If all constraints have been satisfied, then return the solution $\sigma$.
  Otherwise, the constraint set is unsatisfiable.
\end{enumerate}

In the case of function $\syntax{fst}$, the constraints gathered by the rules in
\cref{fig:exp-cgen} are as follows.
  \begin{gather*}
   \kindc{\kindVar_1} \subk {\pKind{\multVar_0}\prekindt},
   \kindc{\kindVar_0} \subk {\pKind{\multVar_0}\prekindt},
   \kindc{\kindVar_1} \subk {\pKind{\multVar_1}\prekindt},
   \kindc{\kindVar_0} \subk {\pKind{\multVar_1}\prekindt},
   \kindc{\kindVar_1} \subk \kindtu,
   \\
   \kindc{\multVar_0} = \multFun{\kindc{\kindVar_0}} \join \multFun{\kindc{\kindVar_1}},
   \kindc{\multVar_1} = \multFun{\kindc{\kindVar_0}} \join \multFun{\kindc{\kindVar_1}}   
     \end{gather*}

 We start with
 $\sigma = [\kind[\kindVar_0]\mapsto\kindtl, \kind[\kindVar_1]\mapsto\kindtl,
 \kindc{\multVar_0}\mapsto\lin, \kind[\multVar_1]\mapsto\lin]$.
 Next we pick
 constraint $\kindc{\kindVar_1} \subk {\pKind{\multVar_0}\prekindt}$ and use
 item 2(a). We have, $\kindc{\kindVar_1} \subk \kindtl$ since
 $\sigma(\kindc{\multVar_0}) = \lin$. Given that $\sigma(\kindc{\kindVar_0})$ is
 equal to $\kindtl$, and subkinding is reflexive, $\sigma(\kindc{\multVar_0})$
 remains as $\kindtl$. The process for the second constraint,
 $ \kindc{\kindVar_1}\subk {\pKind{\multVar_0}\prekindt}$, is similar. We
 analyse the constraint $\kindc{\kindVar_1}\subk\kindtl$ since
 $\sigma(\kindc{\multVar_0}) = \lin$. Also in this case item 2(a) does not
 change $\sigma$. The next two constraints,
 $\kindc{\kindVar_1} \subk {\pKind{\multVar_1}\prekindt}$ and
 $\kindc{\kindVar_0}\subk{\pKind{\multVar_1}\prekindt}$, are also handled by
 item 2(a). Once again, $\sigma$ is subject to no update.
 Now we pick constraint $\kind[\kindVar_1]\subk\kindtu$. Under item 2(a) the
 algorithm computes the greatest lower bound of $\kindtu$ and $\kindtl$, which
 is $\kindtu$, so $\sigma$ is updated accordingly.
 For the last two constraints we use item 2(d). We read the values of
 $\kindc{\kindVar_0}$ and $\kindc{\kindVar_1}$ from $\sigma$ and compute the
 least upper bound of $\multFun\kindtl$ and $\multFun\kindtu$ which yields
 $\lin$. Both entries for $\kindc{\kindVar_0}$ and $\kindc{\kindVar_1}$ are
 already $\lin$ and therefore no update to $\sigma$ is done.
 Since we analysed all constraints and $\sigma$ was updated in this iteration of
 the algorithm, the fixed-point is not reached yet and so we go through each
 constraint once again. This time no update is made and therefore we terminate
 with
 $\sigma = [\kind[\kindVar_0]\mapsto\kindtl, \kind[\kindVar_1]\mapsto\kindtu,
 \kind[\multVar_0]\mapsto\lin, \kind[\multVar_0]\mapsto\lin]$.

 The algorithm iteratively updates the values of the kind and multiplicity
 variables until no further updates can be made, that is, until a fixed point is
 reached. Since the kind lattice is finite, any sequence of updates must
 eventually converge to a fixed point. For the same reason, each constraint can
 only be updated a finite number of times. Therefore, the algorithm terminates
 after a finite number of iterations.

 The running time of the constraint generation algorithm is linear on the size
 of the input expression; that of the constraint satisfaction algorithm is
 quadratic. In the worst case scenario the number of constraints is equal to the
 size of the expression.
Each constraint can only update $\sigma$ twice (when a more restrictive solution
is found). The worst case happens when a different constraint performs an update
in each iteration, forcing the algorithm to analyse all the constraints in each
 iteration.
A sensible optimization removes the constraints from the constraint set also in
items 2(a) and 2(b), after concluding that they cannot update $\sigma$ to a more
restrictive solution. Since the update can only be performed a constant number of
times, the algorithm becomes linear on the size of the input expression.

\paragraph{Evaluation}

\begin{table}[!t]
  \centering
  \begin{tabular}{|l|r|r|}
    \hline 
    Category of annotation
    & \multicolumn{1}{|p{5cm}|}{\centering Number of annotations in the source code}
    & \multicolumn{1}{|p{5cm}|}{\centering Number of more general annotations generated}\\\hline
Datatypes&129&0\\
Type abbreviations&206&7\\
Universal types&282&94\\
Explicit recursive types&23&10\\
Type abstractions&30&25\\\hline
Total&670&136\\ \hline
  \end{tabular}
  \caption{\label{distribution-table}Distribution of annotations}
\end{table}

We implemented the algorithm and incorporated it in the
\freest interpreter. Then we conducted an evaluation to check the
behaviour of the algorithm when used on \freest source code. The evaluation
consisted of replacing all the 670 kind annotations by fresh kind variables in
the 232 valid programs in the \freest test
suite and standard library (total of 9131 lines of code), running the algorithm and checking whether the
algorithm infers the annotations back.

Kind annotations are spread over datatypes, type abbreviations, universal types,
recursive types, and type abstractions. The distribution of annotations is as in
\cref{distribution-table}.
The small number of annotations in recursive types and type abstractions comes
from the fact that they are usually introduced implicitly, either via type
abbreviations (as in the code in line 2) or through compiler
elaboration introducing type abstractions $\ETAbs\TVar\kind\EVal$ for functions
accompanied by their signatures.
%


We concluded that the algorithm correctly inferred all annotations and found
that 136 of the 670 annotations (that is, 20\%) were too specific and could be
relaxed to a more general kind. The largest number of more general annotations
found by the algorithm come from universal types. We attribute this to the
conservative nature of programmers: if we are developing Church encodings (heavy
on polymorphism), why would one require linear type variables? The algorithm did
not improve the kind for datatypes: datatypes are usually used in an
unrestricted manner in programs. Moreover, in the test suite, they usually
appear as the first argument (to be pattern-matched) of functions with
unrestricted closures and therefore they cannot be linear.

For an example where the algorithm suggests a more general kind, consider
function composition.
\begin{lstlisting}
dot : forall a:*T b:*T c:*T . (b -> c) -> (a -> b) -> a -> c
dot f g x = f (g x)
\end{lstlisting}
If we only provide unrestricted arguments to \lstinline|dot|, then there is no
reason why the polymorphic variables \lstinline|a|, \lstinline|b| and
\lstinline|c| could not have kind \lstinline|*T|. However, we would be ruling
out programs that apply \lstinline|dot| to linear arguments. Consider the
following program.

\begin{lstlisting}
dot : (b -> c) -> (a -> b) -> a -> c
dot f g x = f (g x)

g : ?Int;End -> Int
g c = let (x, c) = receive c in close c ; x

main : Int
main =
  let (w,r) = new () in
  fork (\_ 1-> let w = send 5 w in close w);
  dot id g r
\end{lstlisting}
This program would be flagged as untypable because we instantiate the
polymorphic variable \lstinline|a| with the linear session type
\lstinline|?Int;End|. Since there is no reason why \lstinline|a|, \lstinline|b|
and \lstinline|c| should be unrestricted, the algorithm assigns kind
\lstinline|1T| to the three polymorphic variables.

\section{Future Work}


There are several avenues for future work. The most immediate is to prove the
correctness of the algorithm with respect to the typing system.
Then, equipped with kind inference, we may think of introducing a third base
kind, that for session types that must be eventually closed (that reach type
$\TEnd$). In this case we would require the kind of the argument to function
$\ENew$ to be of the newly introduced kind.
We further plan to study the possibility of quantifying over kinds or
multiplicities for extra flexibility in programming.



\begin{sloppypar}
  \paragraph{Acknowledgements}
  We thank the anonymous reviewers for their detailed comments that greatly
  contributed to improve the paper. This work was supported by FCT through
  project SafeSessions, ref.\ PTDC/CCI-COM/6453/2020, and the LASIGE Research
  Unit, ref.\ UIDB/00408/2020 and ref.\ UIDP/00408/2020.
\end{sloppypar}

\nocite{*}
\bibliographystyle{eptcs}
\bibliography{biblio}

\newpage
\appendix


\begin{figure}[t!] 
  \declrel{}{$\cGenK[\Delta_{\text{in}}]{\type_{\text{in}}}{\kind_{\text{out}}}{\cset_{\text{out}}}$}
  \begin{gather*}
      \axiom{\ruleCGUnit}{\cGenK\TUnit{\kind[\pKind\mult\prekindt]}\Empty}
    \quad
      \axiom{\ruleCGVar}
            {\cGenK[\Delta, \TVar\colon\kind] \TVar\kind\Empty}
    \quad        
      \axiom{\ruleCGSkip}{\cGenK \TSkip\kindsu\Empty}            
    \quad
      \axiom{\ruleCGEnd}{\cGenK \TEnd\kindsl\Empty}
    \\
      \infrule{\ruleCGMsg}{\cGenK\type\kind\cset}{\cGenK {\TMsg\type} \kindsl\cset}
    \quad
      \infrule{\ruleCGCh}
              {\cGenK{\type[T_\ell]}{\kind[\kindsym_\ell]}{\cset_\ell} \\ (\forall \ell\in L)}
              {\cGenK{\TChoice{\recordt\ell TL}}{\kindsl}{\bigcup_{\ell\in L} \cset_\ell \cup \{ \kind[\kind_\ell]\subk\kindsl\}}}     
    \\
      \infrule{\ruleCGSeq}
              {\cGenK \type{\kind[\kindsym_1]}{\cset_1} \\
                \cGenK {\type[U]}{\kind[\kindsym_2]}{\cset_2}\\
                \multVar\;\text{fresh}
              }
              {\cGenK {\TSemi\type{\type[U]}}{\pKind\multVar\prekinds}{\cset_1 \cup \cset_2 \cup
                \{{\kind[\kindsym_1]} \subk \kindsl,
                  {\kind[\kindsym_2]} \subk \kindsl,
                  \multVar = \multFun{\kind[\kindsym_1]} \join \multFun{\kind[\kindsym_2]}
                \}}}
    \\
      \infrule{\ruleCGRec}
              {\cGenK [\Delta, \TVar\colon\kind] \type {\kindc{\kind'}} \cset}
              {\cGenK {\TRec a\kind\type} {\kindc{\kind'}}{\cset \cup \{{\kindc{\kind'}} \subk \kind\}}}
     \quad
      \infrule{\ruleCGArrow}
              {\cGenK \type{\kind[\kindsym_1]}{\cset_1} \\
               \cGenK {\type[U]}{\kind[\kindsym_2]}{\cset_2}}
              {\cGenK{\TFun{\type}{\type[U]}}{\pKind\mult\prekindt}{\cset_1 \cup \cset_2}}
    \\        
      \infrule{\ruleCGRcd}
              {\cGenK{\type[T_\ell]}{\kind[\kindsym_\ell]}{\cset_\ell} \\
                \multVar\;\text{fresh}\\
                (\forall \ell\in L)
              }
              {\cGenK{\TRcd\ell TL}{\pKind\multVar\prekindt}
                {\bigcup_{\ell\in L}  \cset_\ell \cup
                  \{\multVar = \bigjoin_{\ell\in L}\multFun{\kind[\kindsym_\ell]}
                  , \kind[\kind_\ell] \subk \pKind\multVar\prekindt
                  \}}}
      \infrule{\ruleCGTAbs}
              {\cGenK[\Delta, \TVar\colon\kind] \type{\kindc{\kind'}}\cset \\\multVar\;\text{fresh}}
              {\cGenK {\TAll a\kind\type} {\pKind\multVar\prekindt} {\cset \cup \{\multVar = \multFun{\kindc{\kind'}}\}}}
  \end{gather*}
  \vspace*{-8mm}
  \caption{Constraint generation from types (complete set of rules)}
  \label{fig:type-contraints-full}
\end{figure}


\begin{figure}[t!] 
  \declrel{}{$\cGenE[\Delta_{\text{in}}]{\Gamma_{\text{in}}}{\expr_{\text{in}}}{\type_{\text{out}}}{\cset_{\text{out}}}{\uset_{\text{out}}}$}
  \begin{gather*}
    \infrule{\rulenameIConst}
            {\cGenK{\typeof\EConst}\kind\cset}
            {\cGenE\Gamma\EConst{\typeof\EConst}\cset\Empty}
    \quad
    \infrule{\rulenameIVar}
            {\cGenK\type\kind\cset}
            {\cGenE{\Gamma,\EVar\colon\type}\EVar\type\cset{\{\EVar\colon\kind\}}}
    \\
            \quad
    \infrule{\rulenameIAbs}
            {\cGenK{\type[T_1]}\kind\cset_1 \\
              \cGenE{\Gamma,\EVar\colon\type[T_1]}\expr{\type[T_2]}{\cset_2}\uset
              \\
              \cset_3 = \isClosure{\expr}{\{\kind\subk{\pKind\mult\prekindt}\}}{\Empty}
            }
            {\cGenE\Gamma{\EAbs x{\type[T_1]}\expr} {\TFun{T_1}{T_2}}
              {\cset_1 \cup \cset_2 \cup \cset_3 \cup \weaken{\EVar}{\kind}{\uset}}
              {\uset\setminus\{\EVar\colon\kind\}}}
            \\
    \infrule{\rulenameIApp}
            {\cGenE\Gamma{\expr[e_1]}{\TFun{T_1}{T_2}}{\cset_1}{\uset_1}
              \\
              \cGenE\Gamma{\expr[e_2]}{\type[T_1]}{\cset_2}{\uset_2}
              \\
              \cGenK{\TFun{T_1}{T_2}}\kind{\cset_3} 
            }
            {\cGenE\Gamma{\EApp {e_1}{e_2}}{\type[T_2]}
              {{\cset_1}\cup{\cset_2}\cup{\cset_3}\cup{\mergeFun{\uset_1}{\uset_2}}}
              {\uset_1 \cup \uset_2}}
            \\
    \infrule{\rulenameITAbs}
            {\cGenE[\Delta,{\TVar}\colon\kind]\Gamma\EVal\type{\cset_1}\uset
              \\
              \cGenK\type{\kind[\kind']}{\cset_2}
            }
            {\cGenE\Gamma{\ETAbs \TVar\kind\EVal}{\TAll a\kind\type}
              {{\cset_1}\cup{\cset_2}}\uset}
            \\
    \infrule{\rulenameITApp}
            {\cGenK\type{\kind[\kindsym_1]}\cset_1 \\
              \cGenE\Gamma\expr{\TAll a{\kind[\kindsym_2]}{\type[U]}}{\cset_2}\uset
            }
            {\cGenE\Gamma{\ETApp\expr\type}{\subs \type \TVar{\type[U]}}
              {{\cset_1}\cup{\cset_2}}\uset}
            \\
   \infrule{\rulenameIRcdElim}
            {\cGenE\Gamma{\expr[e_1]}{\TRcd\ell TL}{\cset_1}{\uset_1}
              \\
              \cGenE{\Gamma,(\expr[x_\ell]\colon{\type[T_\ell]})_{\ell\in L}}
                {\expr[e_2]}\type{\cset_2}{\uset_2}
              \\
              \cGenK{\type}\kind{\cset_3}
              \\
              \cGenK{\type[T_\ell]}{\kind[\kindsym_\ell]}{\cset_\ell}
              \\
              \cset = \cset_1 \cup \cset_2 \cup \cset_3 \cup \mergeFun{\uset_1}{\uset_2} \cup
                \weaken{\expr[x_\ell]}{\kind[\kindsym_\ell]}{\uset_2} 
              \\
              (\forall \ell\in L) 
            }
            {\cGenE\Gamma{\ELet{\ERcd \ell x L}{e_1}{e_2}}\type
              \cset{(\uset_1\cup\uset_2)}\setminus\{ \expr[x_\ell]\colon\kind[\kindsym_\ell] \}_{\ell\in L} }
            \\            
    \infrule{\rulenameIRcd}
            {\cGenE\Gamma{\expr[e_\ell]}{\type[T_\ell]}{\cset_\ell}{\uset_\ell}
              \\
              \cGenK{\type[T_\ell]}{\kindc{\kind_\ell}}\cset'_\ell
              \\
              (\forall \ell\in L)
            }
            {\cGenE\Gamma{\ERcd \ell v L}{\TRcd\ell TL}
              {\cset_\ell \cup \cset'_\ell \cup \merges({\uset_\ell})}
              {\bigcup_{\ell\in L} \uset_\ell}
            }
            \\
    \infrule{\rulenameIVariant}
            { \cGenE\Gamma\expr{\type[T_k]}{\cset_1}\uset
              \\
              \cGenK{\type[T_\ell]}{\kind[\kindsym_\ell]}\cset_\ell
              \\
              k\in L
              \\
              (\forall \ell\in L)
            }
            {\cGenE\Gamma{\EApp k\expr}{\TVariant\ell TL}{\cset_1\cup\cset_\ell}\uset}
            \\
    \infrule{\rulenameICase}
            {\cGenE\Gamma\expr{\TVariant\ell TL}{\cset_1}{\uset_1}
              \\
              \cGenE\Gamma{\expr[e_\ell]}{\TFun{T_\ell}{T}}{\cset_\ell}{\uset_\ell}
              \\
              \cGenK{\type[T_\ell]}{\kind[\kindsym_\ell]}{\cset'_\ell}
              \\
              (\forall \ell\in L)
            }
            {\cGenE\Gamma{\ECase\expr \ell x L}{\type}
              {\cset_1\cup\cset_\ell\cup\cset'_\ell}
              {\uset_1\cup\uset_\ell}}
             \\
   \infrule{\rulenameISel}
           { \cGenK{\type[T_\ell]}{\kind[\kindsym_\ell]}\cset_\ell
             \\
             \cGenE\Gamma{\expr[e_\ell]}{\TFun{T_\ell}T}{\cset'_\ell}{\uset_\ell}
             \\
              k\in L
              \\
              (\forall \ell\in L)
           }
           {\cGenE\Gamma{\EApp \ESel k}{\TFun{\TChoice[\viewOut]{\recordt\ell TL}}
               {\type[T_k]}} {\cset_\ell \cup \cset'_\ell}{\bigcup_{\ell\in L} \uset_\ell}}
           \\
   \infrule{\rulenameINew}
           { \cGenK[\Empty]\type\kind\cset
           }
           {\cGenE\Gamma{\EApp\ENew\type}{\TPair T{\dual T}}\cset\emptyset}
           %
  \end{gather*}
  \caption{Constraint generation from expressions (complete set of rules)}
  \label{fig:exp-cgen-full}
\end{figure}



\end{document}